\documentclass{article}
\usepackage{caption}
\usepackage{subcaption}

\PassOptionsToPackage{numbers, sort}{natbib}

\usepackage[preprint]{neurips_2022}

\usepackage[utf8]{inputenc} 
\usepackage[T1]{fontenc}    
\usepackage{hyperref}       
\usepackage{url}            
\usepackage{booktabs}       
\usepackage{amsfonts}       
\usepackage{nicefrac}       
\usepackage{microtype}      
\usepackage{xcolor}         

\usepackage{multirow}
\usepackage{graphicx}
\usepackage{amsmath,amsfonts,amsthm}
\usepackage{mathtools}
\usepackage{adjustbox}
\usepackage{multicol}
\usepackage{multirow, makecell}
\usepackage[linesnumbered,ruled]{algorithm2e}
\usepackage{amsmath}
\usepackage{wrapfig,lipsum,booktabs}
\usepackage{bbding}

\theoremstyle{plain}

\theoremstyle{definition}

\theoremstyle{remark}

\title{FedAudio: A Federated Learning Benchmark \\ for Audio Tasks}

\author{%
  Tuo Zhang$^{1}$\thanks{The first two authors contribute equally}\footnotemark[1] , TianTian Feng$^{1}$\footnotemark[1] , Samiul Alam$^{2,4}$, \\  
  \textbf{Sunwoo Lee$^{3}$, Mi Zhang$^{4,2}$, Shrikanth S. Narayanan$^{1}$, Salman Avestimehr$^{1}$}
  \\
  \\
  $^{1}$Department of Electrical and Computer Engineering, University of Southern California\\
  $^{2}$Michigan State University\\
  $^{3}$Department of Computer Engineering, Inha University\\
  $^{4}$Department of Computer Science and Engineering, The Ohio State University\\}

\begin{document}

\maketitle

\begin{abstract}
Federated learning (FL) has gained substantial attention in recent years due to the data privacy concerns related to the pervasiveness of consumer devices that continuously collect data from users. 
While a number of FL benchmarks
have been developed to facilitate FL research, none of them
include audio data and audio-related tasks. 
In this paper, we fill this critical gap by introducing a new FL benchmark for audio tasks which we refer to as \texttt{FedAudio}. 
\texttt{FedAudio} includes four representative and commonly used audio datasets from three important audio tasks that are well aligned with FL use cases.
In particular, a unique contribution of \texttt{FedAudio} is the introduction of data noises and label errors to the datasets to emulate challenges when deploying FL systems in real-world settings.
%
\texttt{FedAudio} also includes the benchmark results of the datasets and a PyTorch library with the objective of facilitating researchers to fairly compare their algorithms.
We hope \texttt{FedAudio} could act as a catalyst to inspire new FL research for audio tasks and thus benefit the acoustic and speech research community. The datasets and benchmark results can be accessed at \url{https://github.com/zhang-tuo-pdf/FedAudio}.

\end{abstract}

\vspace{3mm}
\section{Introduction}
\label{sec:intro}

Data privacy has become one of the most critical issues when dealing with personal data. Particularly, audio data, which are now widely accessible in consumer applications like Amazon Alexa, Google Assistant, and Apple Siri, can reveal a significant amount of private information about an individual. 

Recently, federated learning (FL) has emerged as a privacy-preserving solution to address this pressing concern \cite{Zhang2022FederatedLF}.
To facilitate FL research, as summarized in Table~\ref{table:comparison}, a number of FL benchmarks have been developed in the past few years. 
For example, LEAF \cite{Caldas2018LEAFAB} is a FL benchmark that includes five datasets for tasks such as natural language processing (NLP) and computer vision (CV). TensorFlow Federated \cite{Abadi2016TensorFlowAS} expands LEAF by adding three additional datasets from NLP and CV tasks. FedML ecosystem \cite{chaoyanghe2020fedml} introduces several benchmarks for federated learning environment, such as FedGraphGNN \cite{He2021FedGraphNNAF}, FedNLP \cite{Lin2021FedNLPAR}, FedCV \cite{He2021FedCVAF}, and FedIoT \cite{Zhang2021FederatedLF}, targeting Graph Neural Networks (GNN), NLP, CV, Internet of Things (IoT) related applications. Lastly, Flamby \cite{Terrail2022FLambyDA} incorporates seven datasets in the healthcare domain, including both tabular and image data. 
Unfortunately, although these existing benchmarks have made significant contributions on facilitating FL research, 
none of these benchmarks include audio data and audio-related tasks. 

\begin{table}[h]
\caption{Comparing FedAudio with existing FL benchmarks.}
\vspace{1mm}
\small
\centering
\begin{tabular}{lccc}
\toprule  
& \textbf{Data Type} & \textbf{Noisy Data}  & \textbf{Noisy Label} \\
\midrule  
LEAF \cite{Caldas2018LEAFAB} & Image, Text & \XSolid & \XSolid \\
TensorFlow FL \cite{Abadi2016TensorFlowAS} & Image, Text  & \XSolid & \XSolid \\
FedML \cite{chaoyanghe2020fedml} & Graph, Image, Text, Tabular & \XSolid & \XSolid \\
Flamby \cite{Terrail2022FLambyDA} & Image, Tabular & \XSolid & \XSolid \\
\textbf{\textbf{FedAudio}} & \textbf{Audio} & \checkmark & \checkmark \\
\bottomrule 
\end{tabular}
\label{table:comparison}
\end{table}

In this work, we introduce \texttt{FedAudio}, a federated learning benchmark for audio tasks. \texttt{FedAudio} includes four representative and commonly used datasets from three popular audio tasks -- keyword spotting, speech emotion recognition, and sound event classification -- that are well aligned with FL.
Most of the selected datasets are naturally
non-IID distributed by the speaker or actor ID, and include various scales of total client number (from 10 to 2,618).
Unlike existing FL benchmarks listed in Table~\ref{table:comparison} that only include datasets with clean data and accurate labels, in \texttt{FedAudio}, we introduce data noises and label errors to the datasets to emulate scenarios when deploying FL systems in real-world settings. This is a key difference and a unique contribution of \texttt{FedAudio} compared to existing FL benchmarks.
In addition, \texttt{FedAudio} includes the benchmark results of the datasets and a PyTorch library to facilitate researchers to fairly compare their algorithms.
We hope \texttt{FedAudio} could become the reference FL benchmark for audio tasks, and facilitate future FL research in the acoustic and speech research community. 


\vspace{-2.5mm}

\section{Motivation and Design Considerations}
\label{sec:background}

Although FL research has made significant progress over the past few years, the FL research and applications on audio-related tasks are relatively limited. Most FL works on audio-related tasks are conducted either on private datasets \cite{Feng2022FederatedSL, Cui2021FederatedAM} or using different experimental setups \cite{Guliani2021TrainingSR, Dimitriadis2020AFA, Feng2022SemiFedSERSL, Gao2022EndtoEndSR}, making it difficult for researchers to  fairly compare their methods and push the frontier forward. 
Such gap motivates us to develop a high-quality FL benchmark for audio tasks.
to accelerate audio-based FL research.

However, the design of such benchmark requires some unique considerations.
First, the selected datasets need to be both representative and diversified in dimensions such as data size and number of classes, and the selection of the audio tasks needs to align with the use cases of FL. 
%
We do not include the task of speech recognition  mainly because it requires substantial compute resources for large model training and large amount of data labels to achieve competitive accuracy~\cite{Gao2022FederatedSS}.
Second, the benchmark needs to be designed to emulate real-world scenarios as much as possible. As such, the FL algorithms to be developed on top of the benchmark can be more useful when being deployed in real-world settings.
%

\section{FedAudio Design}
\label{sec:fedauido}



\subsection{Datasets}
\label{sec:fedauidoapp}
As listed in Table~\ref{tab:dataset}, \texttt{FedAudio}  includes four audio datasets (Google Speech Commands, IEMOCAP, CREMA-D, Urban Sound) that target three different audio tasks (keyword spotting, emotion recognition, event classification). 
We select these datasets and audio tasks for two primary reasons. 
First, the selected audio tasks are well aligned with FL and are regarded as some of the killer applications of FL. 
Second, the selected datasets are among the most representative and commonly used datasets for the corresponding tasks.
Moreover, the selected datasets cover both small and large numbers of clients, data samples, and class labels, and therefore contribute to a diversified and comprehensive collection.
It should be noted that we do not include the task of automatic speech recognition (ASR). This is mainly because the task of ASR can not be well supported under FL due to its demands on computational resources for large model training and end users for large-amount data labeling \cite{Gao2022FederatedSS}. 

In the following, we describe each dataset along with its pre-possessing method and non-IID data partitioning scheme.

\begin{table}[h]
\tiny
\centering
\caption{
Overview of the datasets included in FedAudio.
}
\vspace{1mm}
\begin{tabular}{ccccrrr} \hline
\toprule
\textbf{Dataset} & \textbf{Task} & \textbf{Pre-processing Method} & \textbf{Non-IID Partition Scheme} & \textbf{\# clients} & \textbf{\# samples} & \textbf{\# labels} \\ 
\midrule
Google Commands \cite{Warden2018SpeechCA} & Keyword Spotting & Mel Spectrogram & Speaker ID & 2,618 & 105,829 & 35\\
\midrule
IEMOCAP \cite{Busso2008IEMOCAPIE} & Speech Emotion Recognition & Pretrained APC model & Actor ID & 10 & 2,943 & 4 \\
\midrule
CREMA-D \cite{cao2014crema} & Speech Emotion Recognition & Pretrained APC model & Speaker ID & 91 & 4,798 & 4\\
\midrule
Urban Sound \cite{Salamon:UrbanSound:ACMMM:14} & Sound Event Classification & Mel Spectrogram & Dirichlet Distribution & 50 & 8,732 & 10\\
\bottomrule
\end{tabular}
\label{tab:dataset}
\end{table}

\subsubsection{Google Speech Commands}
\label{subsec:gcommand}
The Google Speech Commands dataset \cite{Warden2018SpeechCA} is designed for developing basic voice interfaces for the task of keyword spotting. The dataset includes 35 common words from the everyday vocabulary such as "Yes", "No", "Up", and "Down". It contains a total of 105,829 audio recordings collected from 2,618 speakers. The training set includes the recordings from 2,112 speakers and the test set includes the recordings from the rest. 
To pre-process the raw audio data, a sequence of overlapping Hamming windows is applied to the raw speech signal with a time shift of 10 ms. We calculate the discrete Fourier transform (DFT) with a frame length of 1,024 and compute the Mel-spectrogram with a dimension of 128. The Mel-spectrogram is used for training the keyword spotting model.
%
%
The Google Speech Commands dataset is partitioned over speaker IDs, making the dataset naturally non-IID distributed.

\subsubsection{IEMOCAP}
The IEMOCAP dataset \cite{Busso2008IEMOCAPIE} is a multimodal dataset designed for the task of emotion recognition. The dataset contains video, speech, and motion capture of face of emotional expressions collected from ten actors (five males, five females).
%
We extracted the audio component of the IEMOCAP dataset and followed \cite{Zhang2018AttentionBF, Feng2022SemiFedSERSL} to focus on the more challenging improvised sessions and four most frequently occurring emotion labels (neutral, sad, happiness, and anger) for the task of speech emotion recognition (improvised). 
%
We followed \cite{Feng2022SemiFedSERSL} and used the pre-trained autoregressive predictive coding (APC) features \cite{Chung2019AnUA} as the input for training the speech emotion recognition (SER) model.
The dataset is partitioned by the actor ID.

\subsubsection{CREMA-D}
Similar to IEMOCAP, the CREMA-D dataset \cite{cao2014crema} is also a multimodal dataset for emotion recognition.
The corpus includes audio and visual data of utterances spoken under five emotional expressions: happy, sad, anger, fear, and neutral. Similar to the previous work in \cite{Feng2022SemiFedSERSL}, we choose neutral, sad, happiness, and anger emotions as the candidate classification labels.
Compared to IEMOCAP, CREMA-D includes many more actors (91 actors: 48 male, 43 female), and thus can emulate a large-scale client pool in FL. 
Again, we extracted the audio component of the dataset, used the pre-trained APC features for training the SER models, and partitioned the dataset by actor ID.

\subsubsection{Urban Sound}

Urban-Sound \cite{Salamon:UrbanSound:ACMMM:14} is an audio database with 8,732 labeled sound recordings from ten urban sound classes: air conditioner, car horn, children playing, dog bark, drilling, engine idling, gun shot, jackhammer, siren, and street music. 
The audio recordings are spitted into ten folds for training and testing. We further divide each recording into 3-second segments, with an overlap of 0.5 second (in training) and 1 second (in testing) between segments. We extract the Mel-spectrograms using the same setting as described in Section~\ref{subsec:gcommand}. 
%
%
To create non-IID data distributions, we partitioned the dataset into 50 subsets using Dirichlet distribution with $\alpha \in \{0.1, 0.5\}$ to control the level of non-IID where $\alpha = 0.1$ corresponds to high  heterogeneity and $\alpha = 0.5$ corresponds to low  heterogeneity. 

\subsection{New Features}
When deploying FL systems in real-world scenarios, the collected audio data can be interfered with background noises. Moreover, end users may make errors in labeling the audio data, especially for challenging tasks such as emotion recognition. Unlike existing FL benchmarks listed in Table~\ref{table:comparison} that only include datasets with clean data and accurate labels, in \texttt{FedAudio}, we introduce data noises and label errors to the datasets to emulate real-world scenarios. This is a key difference and a unique contribution compared to existing FL benchmarks.

\subsubsection{Data Noises}
Background noises are everywhere in real-world environments.
%
To study the impact of data noises on the performance of FL, we add additive white Gaussian noises (AWGN) to the audio data of the four datasets described in Section~\ref{sec:fedauidoapp}. Specifically, we followed \cite{clavel2005events} to use the signal-to-noise ratio (SNR) as a measurement for data noise level, which is defined as the ratio of the power of the original signal to the power of background noise, and expressed in decibel (dB).


\subsubsection{Label Errors}
Label error refers to the mismatch between the ground truth label of a data sample and the label provided by the end user.
%
To emulate label error, we augment the ground truth labels of a dataset with a transition matrix $Q$, where $Q_{ij} = P(\hat{y}=j\mid y=i)$ denotes the probability that the ground truth label $i$ is changed to a different label $j$.
To do so, we followed \cite{Northcutt2021ConfidentLE} to control the generation of the transition matrix $Q$ using two parameters: error ratio, and error sparsity. 
Specifically, error ratio quantifies the total amount of noises in a dataset, and is defined as the sum of the probabilities that the ground truth label $i$ flips to the other labels. An error ratio of $0$ represents a dataset with all the data accurately labeled, whereas an error ratio of $1$ implies that all the data in a dataset have wrong labels. %
%
On the other hand, error sparsity quantifies the distribution of the label errors, which is defined as the fraction of zeros in the off-diagonal of the transition matrix $Q$. 
%
An error sparsity of $0$ represents each element in $Q$ is non-zero, which implies one class could be altered to any other class. An error sparsity of $1$ indicates that there is no label error as all off-diagonals are equal to zero. 

\subsection{Library}
\begin{figure}[h]
	\centering
	\includegraphics[width=0.55\linewidth]{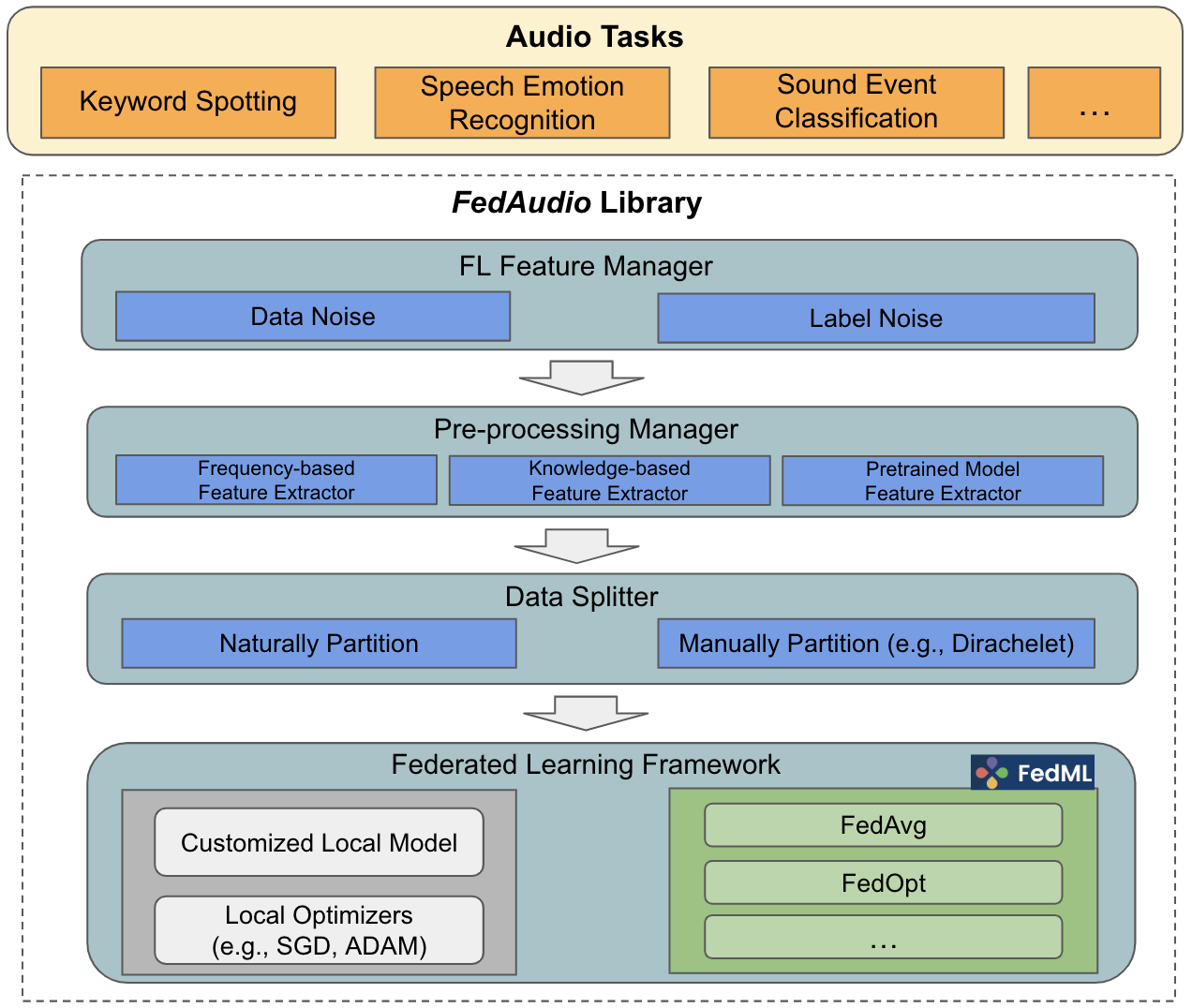}
    \caption{Structure of FedAudio library.}
    \label{fig:fedaudio}
\end{figure}

We have created a library to facilitate the use of \texttt{FedAudio} benchmarks.
As illustrated in Figure~\ref{fig:fedaudio}, our library incorporates a \textit{FL Feature Manager} that adds data and label noise to emulate real-world scenarios. 
The \textit{Pre-processing Manager} supports not only conventional frequency-based features (e.g., Mel-frequency cepstral coefficients (MFCC), Mel Spectrograms) and knowledge-based speech features (e.g., pitch) but also deep audio representations from pre-trained models. 
%
The \textit{Data Spliter} is responsible of partitioning the dataset into non-IID distribution via either natural partitioning (e.g., user ID) or manual partitioning (e.g., Dirichlet distribution). 
Finally, our library supports a handful of commonly used FL optimizers such as FedAvg \cite{McMahan2017CommunicationEfficientLO} and FedOPT \cite{Reddi2021AdaptiveFO}, and allows users to import new models to be trained on the included datasets. In our library, we include FedML open source software~\cite{chaoyanghe2020fedml} (\url{https://github.com/FedML-AI/FedML}) as the FL framework to implement the FedAudio.

\section{Experiments}
\label{sec:result}
\makeatletter
\newcommand\notsotiny{\@setfontsize\notsotiny\@vipt\@viipt}
\makeatother

In this section, we provide benchmark results of the datasets listed in Table~\ref{tab:dataset}. 
To ensure consistency of the results, we used the same model for all the four datasets. Specifically, the model consists of two convolution layers followed by one Gated Recurrent Units (GRU) layer. An average pooling layer is connected to the GRU output, which is then fed through two dense layer to generate the predictions.
%
%
For Google Speech Command, we report the mean and standard deviation of the accuracy over five trials with different random seeds. For the other three datasets, we evaluate each of the train/test splits provided by the datasets and report the mean and standard deviation of F1 score.

The detailed hyperparameter settings are listed in the Appendix~\ref{sec:Appendix}. All experiments are conducted by CPU/GPU simulation. The simulation experiments are conducted on a computing server with eight GPUs. The server is equipped with AMD EPYC 7502 32-Core Processor and 1024G memory. The GPU is NVIDIA RTX A4000.

\subsection{Benchmarks on Clean Audio Data}
First, we provide benchmark results on clean audio data. 
Specifically, we benchmark the performance under different client sample ratios (except the IEMOCAP dataset due to limited number of speakers) and two FL optimizers (FedAvg and FedOPT), and compare it against centralized training. For fair comparison, we use the same local model and optimizer configuration for both centralized and federated training. We aim to benchmark the following performance: (1) What is the performance gap between centralized and federated settings? (2) What is the impact of client sample ratio on the training performance? (3) What is the impact of FL optimizers on the training performance? (4) How does the level of data heterogeneity affect the training performance?

\begin{table}[h]
\small
\begin{center}
\caption{
Benchmarks on clean data. We denote the average and standard deviation of the performance by the notation: avg. (std.)\%. \textbf{\# Samp Cli} stands for the number of sampled clients per round.
}
\vspace{1mm}
\begin{tabular}{cccccc} 
\toprule
\textbf{Dataset} & \textbf{\# Samp Cli} & \textbf{Centralized} &\textbf{FedAvg} & \textbf{FedOPT} \\ 
\midrule
\multirow{4}{*}{\shortstack{Google Command}} & --- & 91.36 (0.30)\% & --- & --- \\
& 106 (5\%) & --- & 88.62 (0.51)\% & 91.05 (0.23)\% \\ 
& 212 (10\%) & --- & 88.44 (0.41)\% & 90.82 (0.09)\% \\ 
& 424 (20\%) & --- & 89.00 (0.24)\% & 90.91 (0.29)\% \\
\midrule

\multirow{2}{*}{IEMOCAP} & --- & 58.95 (4.95)\% & --- & --- \\
& 8 (100\%) & --- & 53.80 (6.61)\% & 55.45 (6.78)\% \\
\midrule

\multirow{4}{*}{CREMA-D} & --- & 70.05 (1.12)\% & --- & --- \\
& 7 (10\%) & --- & 68.80 (2.58)\% & 64.78 (2.94)\% \\ 
& 21 (30\%) & --- & 68.93 (3.34)\% & 66.51 (2.30)\% \\ 
& 36 (50\%) & --- & 67.63 (3.87)\% & 64.93 (4.33)\% \\
\midrule

\multirow{3}{*}{\shortstack{Urban Sound\\($\alpha=0.5$)} } & --- & 62.83 (4.17)\% & --- & --- \\
& 10 (20\%) & --- & 59.23 (4.21)\% & 57.53 (5.46)\% \\ 
& 25 (50\%) & --- & 59.52 (4.84)\% & 61.16 (7.72)\% \\ 
\cmidrule(lr){2-5}
\multirow{3}{*}{\shortstack{Urban Sound\\($\alpha=0.1$)} } & --- & 62.83 (4.17)\% & --- & --- \\
& 10 (20\%) & --- & 52.49 (7.34)\% & 51.13 (5.46)\% \\ 
& 25 (50\%) & --- & 54.01 (6.87)\% & 56.07 (7.84)\% \\ 

\bottomrule
\label{tab:clean}
\end{tabular}
\end{center}
\end{table}

\vspace{2mm}
\noindent
\textbf{Benchmark Results:}
Table~\ref{tab:clean} summarizes our results. 
(1) The gaps between centralized and FL settings across all datasets are between 0.31\% to 11.7\%. For the same speech emotion recognition task, IEMOCAP achieves lower accuracy than CREMA-D. One plausible reason is that data included in IEMOCAP are from actors' improvisation. In contrast, the data from CREMA-D are strictly based on the script reading, which provides less variation compared to IEMOCAP. 
(2) The increasing of client sample ratio does not necessarily bring performance gains. 
(3) Both FL optimizers achieve competitive performance. 
(4) There is a gap in training performance between high and low data heterogeneity and higher data heterogeneity has a negative effect on the training performance.

\subsection{Benchmarks on Noisy Audio Data}
Next, we provide benchmark results on noisy audio data.
Besides model accuracy/F1 score, we also use the round-to-accuracy metric, which is defined as the  communication rounds needed for training the model to reach a target accuracy. 
We use the smallest client sample ratio combined with the FedAvg optimizer from Table~\ref{tab:clean} as the baseline. 
For Urban Sound, we consider the more challenging high data heterogeneity setting.
$\Delta$Metric denotes the model accuracy/F1 score difference between the baseline and the results on noisy audio data. 
We aim to benchmark the following performance: (1) What is the impact of data noises on model accuracy/F1 score and round-to-accuracy? (2) What is the impact of SNR on training performance?

\begin{table}[t]
\small
\begin{center}
\caption{
Benchmarks on data with noises. “$>$” means the target accuracy was not reached.
}
\vspace{1mm}
\begin{tabular}{cccccc} 
\toprule
\textbf{Dataset} & \textbf{SNR} & \textbf{$\Delta$Metric} &\textbf{Target Acc.} & \textbf{Training Rounds} \\ 
\midrule
\multirow{6}{*}{\shortstack{Google Command}} & 
\multirow{2}{*}{10dB} & 
\multirow{2}{*}{$\downarrow$27.12 (2.11)\%}
& 65\% & $>5000$ \\
&&& 75\% & $>5000$ \\
\cmidrule(lr){2-5}
&\multirow{2}{*}{20dB} & 
\multirow{2}{*}{$\downarrow$12.74 (0.64)\%}
& 65\% & 910 (1.52×) \\
&&& 75\% & 2110 (2.32×) \\
\cmidrule(lr){2-5}
&\multirow{2}{*}{30dB} & 
\multirow{2}{*}{$\downarrow$4.48 (0.39)\%}
& 65\% & 775 (1.29×) \\
&&& 75\% & 1485 (1.63×) \\
\midrule

\multirow{6}{*}{IEMOCAP} & 
\multirow{2}{*}{10dB} & 
\multirow{2}{*}{$\downarrow$5.99 (1.90)\%}
& 35\% & 66 (1.32×) \\
&&& 45\% & 142 (1.65×) \\
\cmidrule(lr){2-5}
&\multirow{2}{*}{20dB} & 
\multirow{2}{*}{$\downarrow$5.56 (2.32)\%}
& 35\% & 59 (1.18×) \\
&&& 45\% & 111 (1.29×) \\
\cmidrule(lr){2-5}
&\multirow{2}{*}{30dB} & 
\multirow{2}{*}{$\downarrow$4.82 (2.78)\%}
& 35\% & 55 (1.10×) \\
&&& 45\% & 103 (1.20×) \\
\midrule
\multirow{6}{*}{CREMA-D} & 
\multirow{2}{*}{10dB} & 
\multirow{2}{*}{$\downarrow$9.11 (2.24)\%}
& 55\% & 58 (1.45×) \\
&&& 65\% & $>200$ \\
\cmidrule(lr){2-5}
&\multirow{2}{*}{20dB} & 
\multirow{2}{*}{$\downarrow$4.43 (2.40)\%}
& 55\% & 47 (1.18×) \\
&&& 65\% & 149 (1.73×) \\
\cmidrule(lr){2-5}
&\multirow{2}{*}{30dB} & 
\multirow{2}{*}{$\downarrow$1.58 (2.40)\%}
& 55\% & 42 (1.05×) \\
&&& 65\% & 128 (1.48×) \\
\midrule

\multirow{6}{*}{\shortstack{Urban Sound\\($\alpha=0.1$)}} & 
\multirow{2}{*}{10dB} & 
\multirow{2}{*}{$\downarrow$28.07 (3.43)\%}
& 20\% & 184 (3.93×) \\
&&& 25\% & $>300$ \\
\cmidrule(lr){2-5}
&\multirow{2}{*}{20dB} & 
\multirow{2}{*}{$\downarrow$25.51 (4.23)\%}
& 20\% & 145 (3.09×) \\
&&& 25\% & 211 (3.44×) \\
\cmidrule(lr){2-5}
&\multirow{2}{*}{30dB} & 
\multirow{2}{*}{$\downarrow$24.59 (4.43)\%}
& 20\% & 157 (3.56×) \\
&&& 25\% & 176 (2.87×) \\
\bottomrule
\label{tab:snr}
\end{tabular}
\end{center}
\end{table}

\vspace{2mm}
\noindent
\textbf{Benchmark Results:}
Table~\ref{tab:snr} summarizes our results.
(1) AWGN not only degrades model accuracy/F1 score but also increases the required training rounds to the target performance. One notable point is that the AWGN significantly degrades the model performance for Urban Sound, which reveals the vulnerability of sound event classification to the white background noise.
(2) With the increment of SNR level, we find improvements in both model accuracy/F1 score and round-to-accuracy. These results indicate that noise cancellation is necessary before feeding the audio data to the FL algorithm. However, even with $30\text{dB}$ SNR, although the model performance decreases by a small margin, the convergence speed is delayed, which brings attention to developing more robust FL strategies on the system side under these settings.


\vspace{-0mm}
\begin{table}[h]
\small
\begin{center}
\caption{
Benchmarks on data with label errors. “$>$” means the target accuracy was not reached.\\
\textbf{E.R.} stands for error ratio.
}
\vspace{1mm}
\begin{tabular}{cccccc} 
\toprule
\textbf{Dataset} & \textbf{E.R.} & \textbf{$\Delta$Metric} & \textbf{Target Acc.} & \textbf{Training Rounds} \\ 
\midrule
\multirow{6}{*}{\shortstack{Google Command}} & 
\multirow{2}{*}{0.1} & 
\multirow{2}{*}{$\downarrow$0.38 (0.16)\%}
& 75\% & 1050 (1.75×) \\
&&& 85\% & 2450 (2.69×) \\
\cmidrule(lr){2-5}
&\multirow{2}{*}{0.3} & 
\multirow{2}{*}{$\downarrow$2.23 (0.25)\%}
& 75\% & 1460 (2.43×) \\
&&& 85\% & 3710 (4.08×) \\
\cmidrule(lr){2-5}
&\multirow{2}{*}{0.5} & 
\multirow{2}{*}{$\downarrow$5.25 (0.28)\%}
& 75\% & 2335 (3.89×) \\
&&& 85\% & $>5000$ \\
\midrule

\multirow{6}{*}{IEMOCAP} & 
\multirow{2}{*}{0.1} & 
\multirow{2}{*}{$\downarrow$4.38 (4.10)\%}
& 35\% & 56 (1.13×) \\
&&& 45\% & 106 (1.24×) \\
\cmidrule(lr){2-5}
&\multirow{2}{*}{0.3} & 
\multirow{2}{*}{$\downarrow$9.64 (1.91)\%}
& 35\% & 80 (1.61×) \\
&&& 45\% & $>200$ \\
\cmidrule(lr){2-5}
&\multirow{2}{*}{0.5} & 
\multirow{2}{*}{$\downarrow$19.43 (6.53)\%}
& 35\% & $>200$ \\
&&& 45\% & $>200$ \\
\midrule

\multirow{6}{*}{CREMA-D} & 
\multirow{2}{*}{0.1} & 
\multirow{2}{*}{$\downarrow$0.39 (2.85)\%}
& 55\% & 51 (1.28×) \\
&&& 65\% & 112 (1.30×) \\
\cmidrule(lr){2-5}
&\multirow{2}{*}{0.3} & 
\multirow{2}{*}{$\downarrow$4.68 (1.16)\%}
& 55\% & 84 (2.00×) \\
&&& 65\% & $>200$ \\
\cmidrule(lr){2-5}
&\multirow{2}{*}{0.5} & 
\multirow{2}{*}{$\downarrow$8.68 (3.76)\%}
& 55\% & 134 (3.36×) \\
&&& 65\% & $>200$ \\
\midrule

\multirow{6}{*}{\shortstack{Urban Sound\\($\alpha=0.1$)}} & 
\multirow{2}{*}{0.1} & 
\multirow{2}{*}{$\downarrow$0.71 (7.12)\%}
& 35\% & 101 (1.21×) \\
&&& 45\% & 186 (1.32×) \\
\cmidrule(lr){2-5}
&\multirow{2}{*}{0.3} & 
\multirow{2}{*}{$\downarrow$4.62 (7.49)\%}
& 35\% & 153 (1.83×) \\
&&& 45\% & 211 (1.50×) \\
\cmidrule(lr){2-5}
&\multirow{2}{*}{0.5} & 
\multirow{2}{*}{$\downarrow$10.98 (7.41)\%}
& 35\% & 221 (2.66×) \\
&&& 45\% & $>300$ \\
\bottomrule
\label{tab:label}
\end{tabular}
\end{center}
\end{table}

\subsection{Benchmarks on Audio Data with Label Errors}
\vspace{-1mm}
Finally, we provide benchmark results on audio data with label errors.
%
For each dataset, we set the error ratio in the range of 0.1 to 0.5 with a step size of 0.2, and set the error sparsity to 0.4 to emulate non-IID distribution.
Again, for Urban Sound, we consider the high data heterogeneity setting.
We aim to benchmark the impact of label errors on model accuracy/F1 score as well as  round-to-accuracy.

\vspace{2mm}
\noindent
\textbf{Benchmark Results:}
Table~\ref{tab:label} summarizes our results.
Among all the datasets, with the error ratio below 0.3, the model performance remains relatively high, but the learning process takes much more compared to both clean and noisy data. When the error ratio reaches 0.5, the model performance drops substantially. The results imply that as the amount of correctly labeled data samples is above a specific threshold, the model could still be trained robustly but with slow convergence speed and a slight decrease in the model performance. We can also find that each dataset has a different performance drop at an error ratio of 0.5. For example, the final model performance drop is 5.19\% in Google Command but is 19.43\% in IEMOCAP. Since the baseline Google Speech Command performance is much higher than IEMOCAP, we hypothesize that this difference is introduced by the task characteristics, where the model performance of a relatively more challenging speech task is less prone to label noises.


\vspace{0mm}
\section{Conclusion}
\vspace{0mm}
We introduced a federated learning benchmark for audio tasks named \texttt{FedAudio}. 
Currently, \texttt{FedAudio} includes four representative datasets of three important audio tasks. 
Besides the clean audio data, our paper investigates how real-world challenges, including data noises and label errors, affect the FL training performance. 
We would continue enriching the package with datasets and baseline from other applications to support more research scenarios.
We hope \texttt{FedAudio} could act as a catalyst to inspire new FL research in the acoustic and speech research community. 

\newpage
\bibliographystyle{plain}
\bibliography{mybib}



\appendix

\newpage
\section{Appendix}
\label{sec:Appendix}

\subsection{Hyperparameter Settings}
We searched for the client learning rate in a range from $10^{-6}$ to $10^0$, the server learning rate in a range from $10^{-4}$ to $10^0$, and input batch size in a range from $5$ to $30$, and total
training round in a range from $50$ to $5000$. After hyperparameter searching, we fixed the batch size for all datasets to 16 and the local epoch to $1$ for all experiments. 
\subsubsection{Training with clean signals}
When we train the model with clean signals, we select both FedAvg and FedOpt as the server aggregator functions. For FedOpt, we select the ADAM as the server optimizer. The other hyperparameters are shown in Table~\ref{tab:hyper1}.

\begin{table}[h]
\small
\begin{center}
\caption{
Hyperparameters for training FL models with clean audio signals.
}
\begin{tabular}{crrrrr} \hline
\toprule
Dataset & Agg. & Sample rate & lr & Server lr & Round \\ \hline
\midrule
\multirow{6}{*}{Google Command} 
&\multirow{3}{*}{FedAvg} & 5\% & 0.1 & - & 5000 \\
&& 10\% & 0.3 & - & 5000 \\
&& 20\% & 0.2 & - & 5000 \\
\cmidrule(lr){2-6}
&\multirow{3}{*}{FedOpt} & 5\% & 0.05 & 0.001 & 5000 \\
&& 10\% & 0.01 & 0.001 & 5000 \\
&& 20\% & 0.01 & 0.001 & 5000 \\
\hline
\midrule

\multirow{2}{*}{IEMOCAP} 
&FedAvg & 100\% & 0.01 & - & 200 \\
\cmidrule(lr){2-6}
&FedOpt & 100\% & 0.01 & 0.001 & 50 \\
\hline
\midrule

\multirow{6}{*}{Crema-D} 
&\multirow{3}{*}{FedAvg} & 10\% & 0.1 & - & 200 \\
&& 30\% & 0.1 & - & 200 \\
&& 50\% & 0.1 & - & 200 \\
\cmidrule(lr){2-6}
&\multirow{3}{*}{FedOpt} & 10\% & 0.1 & 0.001 & 200 \\
&& 30\% & 0.1 & 0.001 & 200 \\
&& 50\% & 0.1 & 0.001 & 200 \\
\hline
\midrule

\multirow{4}{*}{Urban Sound} 
&\multirow{2}{*}{FedAvg} & 20\% & 0.075 & - & 300 \\
&& 50\% & 0.075 & - & 300 \\
\cmidrule(lr){2-6}
&\multirow{2}{*}{FedOpt} & 20\% & 0.1 & 0.001 & 300 \\
&& 50\% & 0.1 & 0.001 & 300 \\
\hline
\bottomrule
\label{tab:hyper1}
\end{tabular}
\end{center}
\end{table}

\subsubsection{Training with SNR noisy signals}
When we train the model with SNR noisy signals, we select FedAvg as the server aggregator function. For Google command related experiments, the client learning rate is 0.1 and the communication round is 5000. For IEMOCAP related experiments, the client learning rate is 0.01 and the communication round is 200. For CREMA-D related experiments, the client learning rate is 0.1 and the communication round is 200. Finally, for Urban Sound, the client learning rate is 0.075 and the communication round is 300.

\subsubsection{Training with label noisy signals}
In this section, we select FedAvg as the server aggregator function. For Google command related experiments, the client learning rate is 0.1 and the communication round is 5000. For IEMOCAP related experiments, the client learning rate is 0.01 and the communication round is 200. For CREMA-D related experiments, the client learning rate is 0.1 and the communication round is 200. Finally, for Urban Sound, the client learning rate is 0.075, and the communication round is 300.

\subsection{Experiments with both data noise and label errors}
\begin{figure*}[h]
	\centering
	\includegraphics[width=0.85\linewidth]{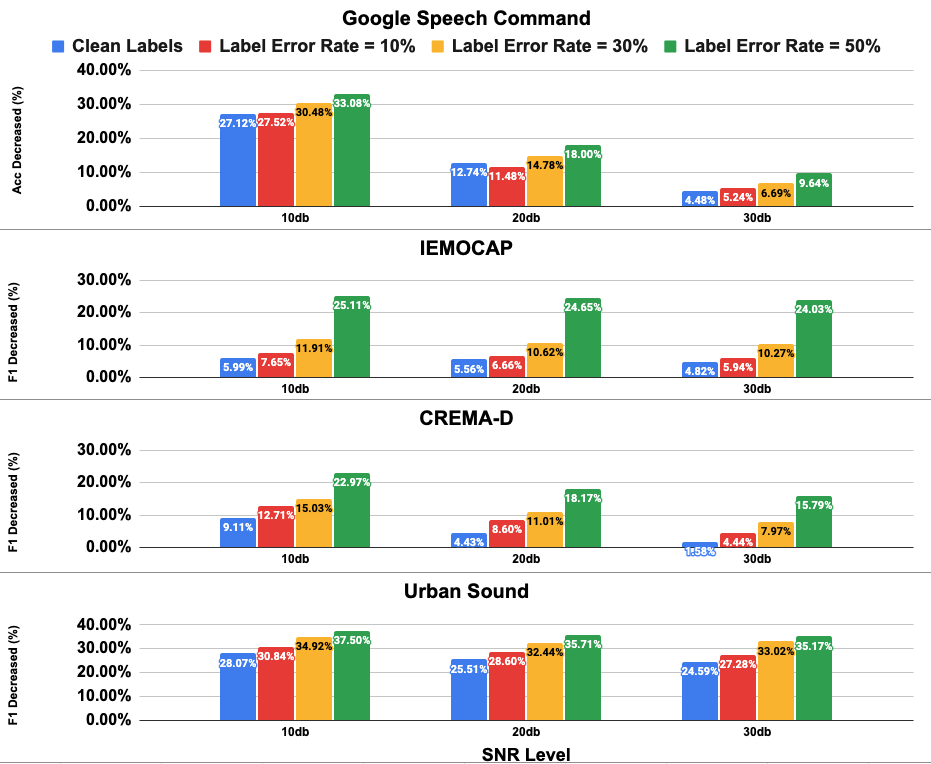}
    \caption{Evaluation results with both data noise and label error.}
    \label{fig:two_noise}
    \vspace{-2.5mm}
\end{figure*}

Besides the noise from a single source, we also conduct the experiments with both data noise and label errors to see the impact of the combination. Figure~\ref{fig:two_noise} shows the model accuracy/F1 score difference between the baseline and the results on noisy audio data. As expected, the combination of data noise and label error has a more negative impact compared to single noise source across all four datasets.

\end{document}